\documentclass[3p,number,preprint,times]{elsarticle} 

\usepackage[utf8]{inputenc}
\usepackage[T1]{fontenc}
\usepackage{amsmath}
\usepackage{amsfonts}
\usepackage{amssymb}
\usepackage{graphicx}
\usepackage{hyperref}

\usepackage[left]{lineno}

\begin{document}

\begin{abstract}
We present a comprehensive characterization of the LAPPD Gen‑II, a large-area microchannel-plate photomultiplier tube (MCP-PMT) equipped with a capacitively coupled sensing (readout) electrode. Two detector variants with different geometries and materials were investigated using a picosecond pulsed laser system. We measured the single-photon timing response and spatial charge distribution on segmented readout electrodes. The prompt timing peak exhibits a resolution of approximately 30~ps, with the overall timing structure explained by photoelectron propagation and back-scattering effects from the MCP input surface. We developed analytical models that describe the propagation of photoelectrons and the induced charge spread on the sensing electrodes, including secondary electron backscattering from the resistive anode. The model accurately reproduces the measured device properties and enables performance extrapolations for various detector geometries and dielectric properties. These results provide a predictive framework for optimizing MCP-PMTs for timing- and imaging-critical applications such as RICH detectors in high-energy physics and TOF-PET systems for medical imaging.
\end{abstract}

\begin{frontmatter}
\title{Characterisation of the LAPPD, a large area microchannel-plate PMT}
\author[UM,IJS]{S. Korpar} 
\cortext[cor1]{Corresponding author}
\ead{Samo.Korpar@um.si}

\author[ULJ,IJS]{R. Dolenec}
\author[ULJ]{F. Grijalva}
\author[IJS]{A. Lozar}
\author[ULJ]{A. Kodri\v c}  
\author[ULJ,IJS]{P. Kri\v zan} 
\author[IJS]{S. Parashari\footnote{Present address: IFIC, Valencia, Spain}}
\author[IJS]{R. Pestotnik}
\author[IJS]{A. Seljak} 
\author[IJS]{D. \v Zontar} 

\address[UM]{Faculty of Chemistry and Chemical Engineering, University of Maribor,
Maribor, Slovenia}
\address[IJS]{Jo\v zef Stefan Institute, Ljubljana, Slovenia}
\address[ULJ]{Faculty of Mathematics and Physics, University of Ljubljana, Ljubljana, Slovenia}

\date{November 2025}
\end{frontmatter}

\section{Introduction}
Cutting-edge photosensor technology plays a crucial role in the evolving landscape of particle physics and medical instrumentation, with applications such as Ring Imaging Cherenkov (RICH) detectors and Positron Emission Tomography (PET) systems. Demanding requirements — including picosecond-level time resolution, high spatial precision, MHz-level counting capability, low dark-count rates, minimal sensitivity to magnetic fields, radiation tolerance, and cost efficiency — are driving the next generation of technological innovation~\cite{DRD4,LHCb:2021glh,Belle-II:2022cgf,lecoq10ps,razdevsek-panel-tofpet}.

Photomultiplier tubes with microchannel plates (MCP-PMT) are well-known low-light level sensors with excellent timing properties~\cite{Akatsu:2004mq,Vavra:2006kdt,korpar-burle-3,mcppmt-timing-tremsin,lappd-timing-adams,DOLENEC2024169864,Lehmann:2024grg}.  Large Area Picosecond Photodetector (LAPPD)~\cite{lappd,lappd-production} is a novel multichannel MCP-PMT. It provides single-photon sensitivity, sub-millimeter spatial resolution~\cite{Kiselev:2021ypx}, temporal resolution of a few tens of picoseconds~\cite{lappd-timing-adams,DOLENEC2024169864}, and a large active area (around $20 \times 20$~cm$^2$). 

An advanced version of the LAPPD, referred to as Gen-II, features a resistive anode that is capacitively coupled through a back plate at the bottom of the detector to the sensing electrode located outside the vacuum region (Fig.~\ref{fig:layout}). This design broadens the range of feasible readout systems and layouts while simplifying the construction of the vacuumized detector volume.

In this article, we present performance studies of this new version of LAPPD. Our focus is on understanding how capacitive coupling influences the spatial distribution of signals on the sensing electrode and the achievable time resolution, building on the model developed in our previous work~\cite{Korpar:2008ria,korpar-burle-3}. An additional objective of this study is to determine the optimal set of MCP-PMT parameters for different application domains, particularly RICH and PET systems, each of which imposes distinct performance requirements.

In what follows, we will present the photo-sensor, describe the experimental setup, present the measurement results, discuss the modeling of effects in the device, and draw our conclusions.

\section{MCP-PMTs investigated in this study}

\begin{figure}
\centering 
\includegraphics[width=0.55\columnwidth]{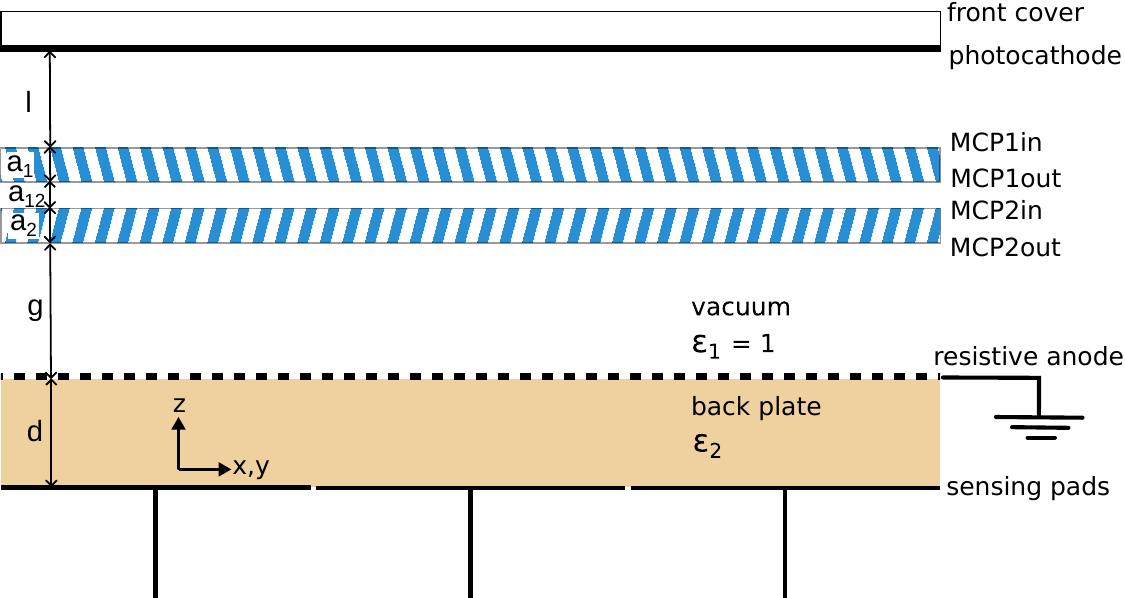}
\includegraphics[width=0.4\columnwidth]{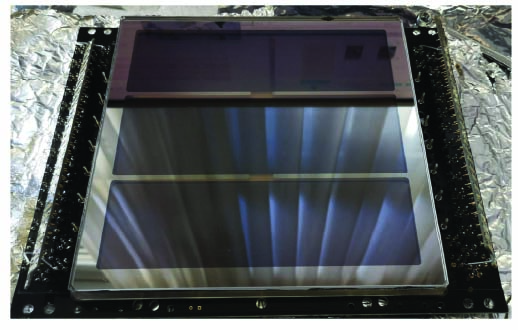}
\caption{LAPPD sensor cross-section: schematic layout of the LAPPD detector with a definition of the main potential levels, geometric parameters, and dielectric constants (left); photo of the front face of the LAPPD detector (right).}
\label{fig:layout}
\end{figure}

\begin{figure}
\centering 
\includegraphics[width=0.50\columnwidth]{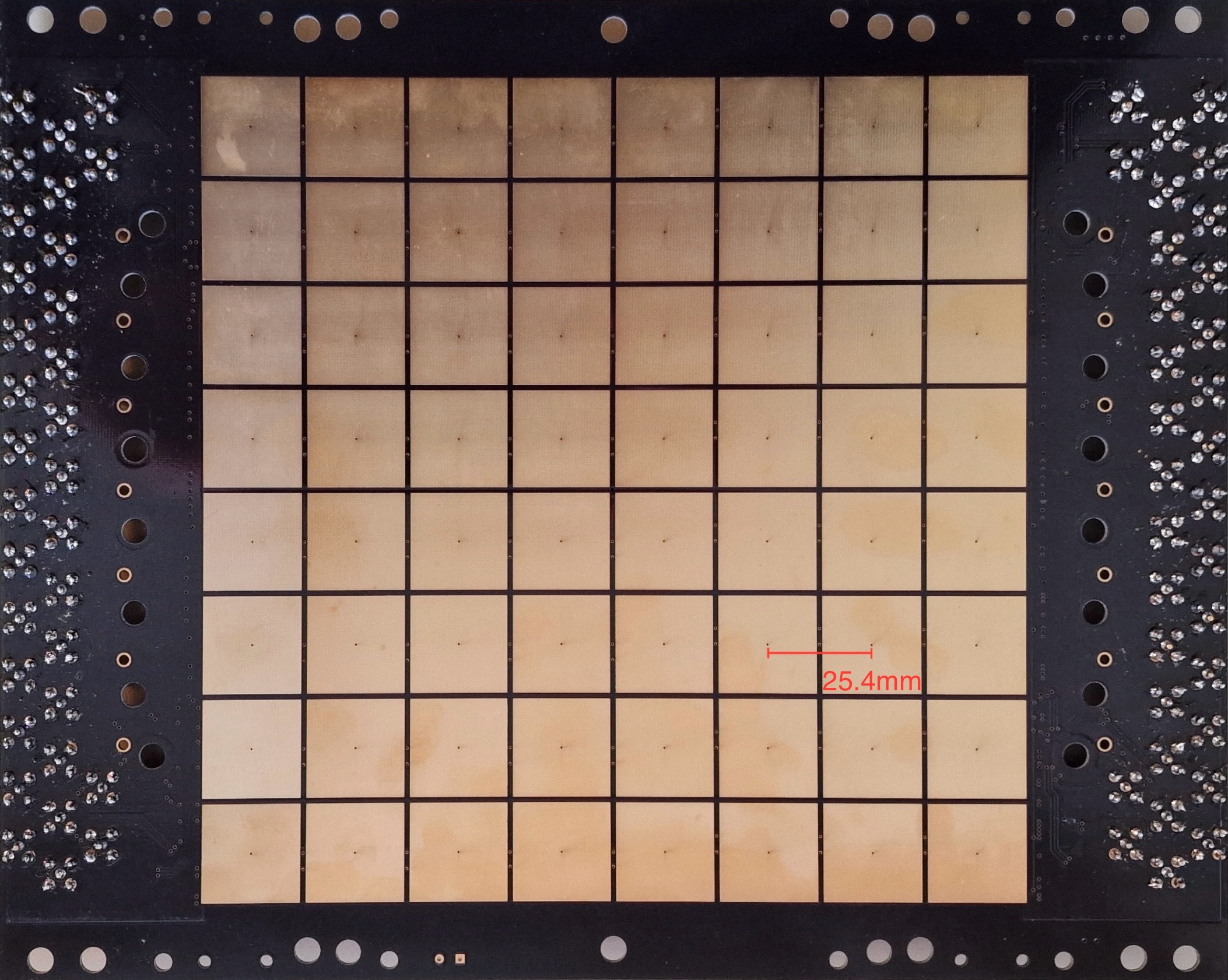}
\caption{External read-out electrode, the original (Incom) sensing electrode with pads at 25.4~mm pitch.}
\label{fig:readout}
\end{figure}

The detectors used in this study were LAPPD Gen-II detectors developed by Incom Inc.~\cite{lappd}.  They use a front cover with a 5~mm thick fused silica glass window and multi-alkali photocathode for photon conversion, two MCPs stacked in a chevron pair for signal multiplication, and a back plate with an interior monolithic resistive anode that is capacitively coupled to the readout electrode on the external side (Fig.~\ref{fig:layout}). The MCPs of the ALD-GCA type are produced from Incom C5 Glass capillary arrays (GCA) and covered by atomic layer deposition (ALD) with layers of resistive and emissive coatings ~\cite{lappd-production}. 
\begin{table}
\caption{LAPPD MCP-PMT parameters.}
\label{tab:mcp-pmts}
\begin{center}
\begin{tabular}{|l|c|c|c|c|c|}
\hline
serial & photocathode-to-MCP1& MCP2-to-anode & resistive anode & back plate material and& microchannel  \\
number & distance $l$ & distance $g$ & thickness $d$ &  dielectric constant  $\epsilon_2$  & diameter\\
\hline
\#162 & 1.19~mm & 3~mm &  2~mm & ceramics, 10 & 10~$\mu$m \\
\hline
\#109 & 2.8~mm & 6.7~mm & 5~mm & borosilicate glass, 4.6 & 20~$\mu$m\\
\hline
\end{tabular}
\end{center}
\end{table}
The parameters of the two samples are listed in Table~\ref{tab:mcp-pmts}. The more recent sample, Incom \#162, has smaller gaps between the photocathode and the first MCP and between the second MCP and the resistive anode; it also has a thinner back plate made of a material with a higher dielectric constant (ceramic instead of borosilicate glass). The multi-alkali photocathodes are Na$_2$KSb and K$_2$NaSb for the tubes \#162 and \#109, respectively.   

Both samples enable independent control of voltages to the photocathode and MCPs.  The original (Incom) sensing electrode with square sensing pads at 25.4~mm pitch (Fig.~\ref{fig:readout}) was used to read-out the signals from the MCP-PMT. 
Signals from the sensing electrode were transmitted through inductors to the SMA connectors, ensuring a positive polarity when connected to the read-out system.

\section{Experimental set-up}

\begin{figure}
\centering 
\includegraphics[width=0.5\columnwidth]{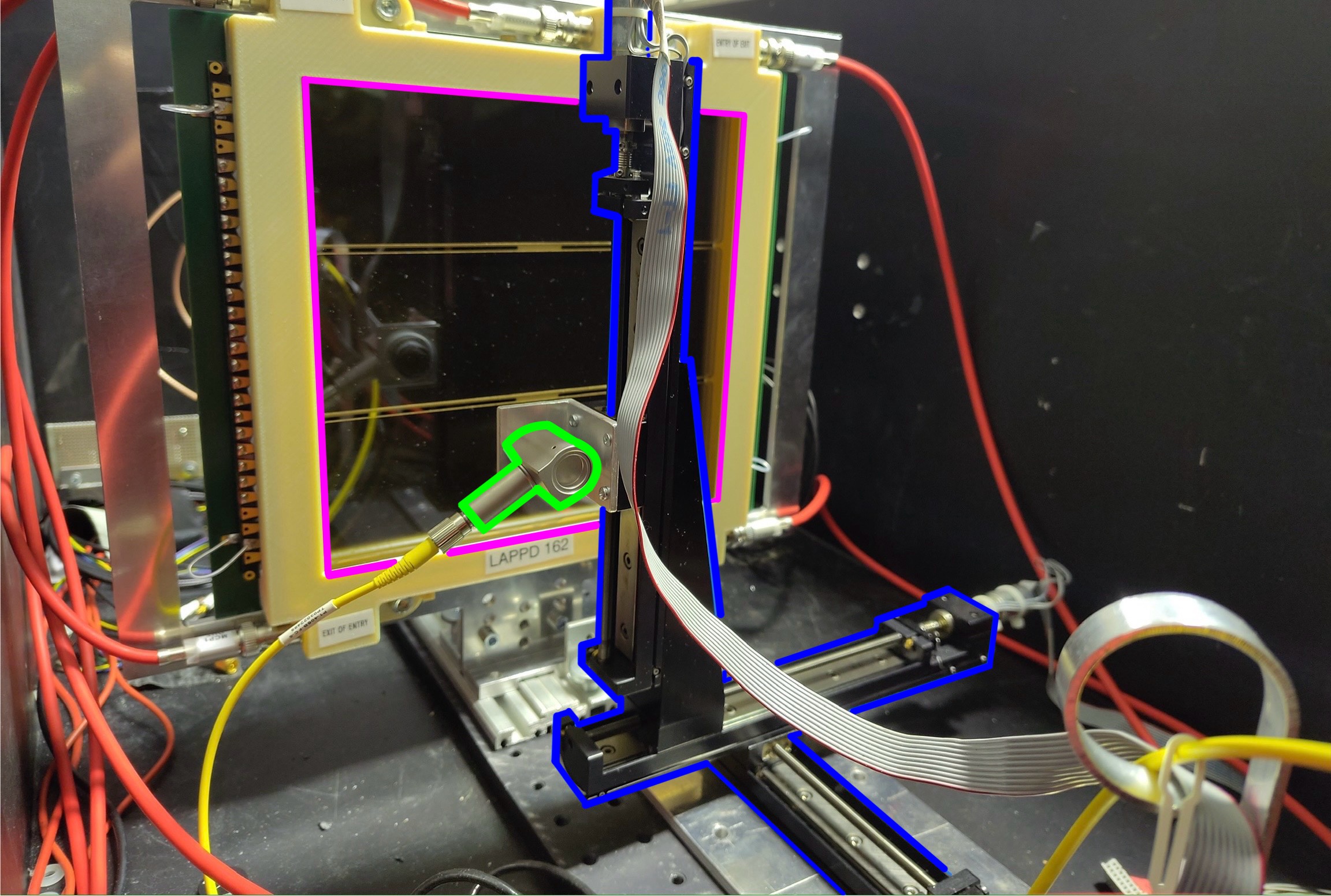}
\caption{The LAPPD detector in the measurement setup; the laser light propagates through the yellow fibre to the focuser (outlined in green) that moves with the 3D stages (outlined in blue) to the photocathode window (outlined in magenta). The red cables connect the MCPs to the power supply for biasing, and the ribbon cable connects to the Z stage.} 
\label{fig:setup}
\end{figure}

The measurement setup (Fig.~\ref{fig:setup}) used an ALPHALAS PICOPOWER™-LD diode laser~\cite{alphalas} that emitted light pulses (20~ps FWHM) with a wavelength of 405~nm focused on a spot with a diameter of approximately 10~$\mu$m. The laser output was attenuated to provide mostly single photons per pulse on the detector and led by an optical fiber into a dark box with the LAPPD. 
A computer-controlled system was used to move the light spot across the detector. 

The five high-voltage levels required to power the LAPPD 
(Fig.~\ref{fig:layout}) were provided by a CAEN HiVolta (DT1415ET) power supply. The standard settings for sample \#109 were 
$U_{\rm PC-MCP1in}=100$~V, 
$U_{\rm MCP1in-MCP1out}=825$~V,
$U_{\rm MCP1out-MCP2in}=200$~V, 
$U_{\rm MCP2in-MCP2out}=825$~V and
$U_{\rm MCP2out-A}=500$~V; for \#162, the corresponding values were: 50~V, 850~V, 200~V, 850~V, and 200~V, respectively. 
The signals from the sensing electrode pads were amplified (ORTEC FTA 820B), discriminated (Phillips Scientific 
discriminator model 806), and timed with a CAMAC TDC (KaizuWorks KC3781A) 
for which the laser controller provided the stop pulse. The 
analog pulse was measured by a VME charge integrator (CAEN QDC V965), 
and the entire measurement was controlled by a PC running LabWindows. 

\begin{figure}
\centering 
\includegraphics[width=0.6\columnwidth]{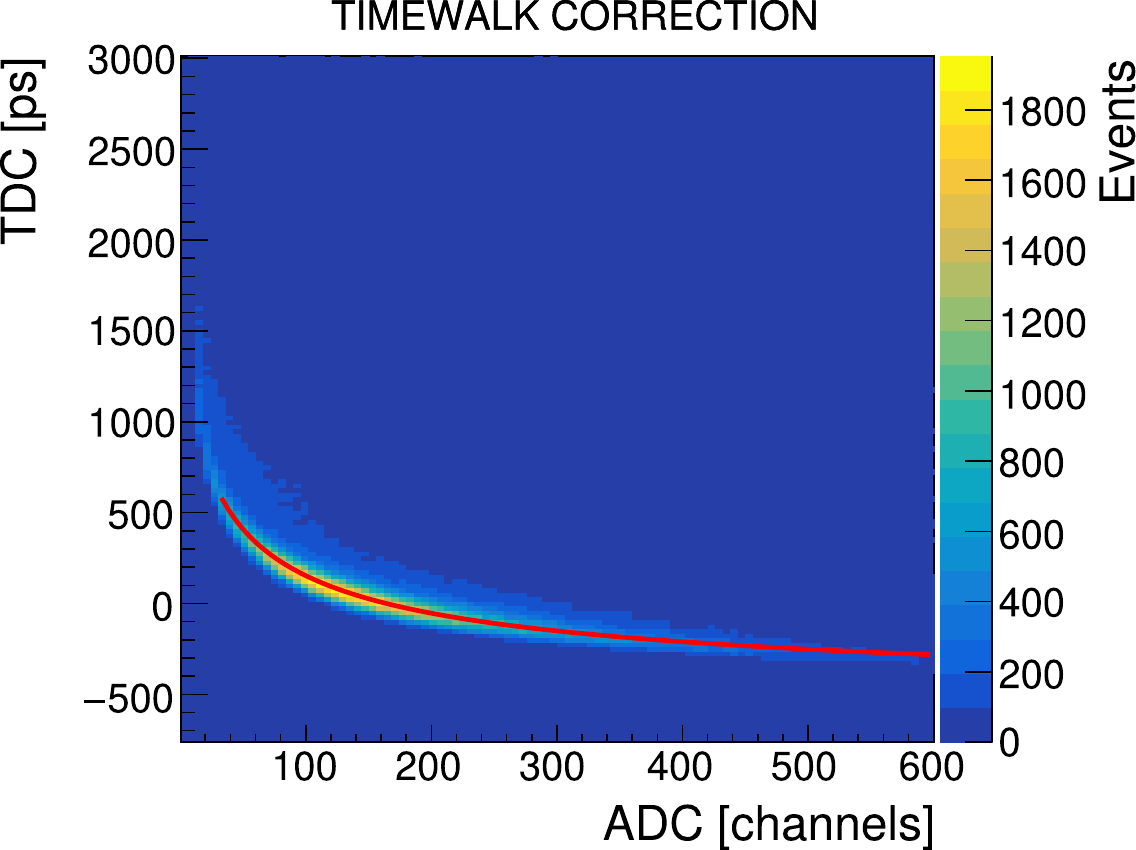}
\caption{Time-walk correction: signal timing as a function of signal charge 
raw data with the correction curve (in red,     $f(ADC)=-612\>\rm{ps} + 8166\>\rm{ps}/\sqrt{ADC\left[channels\right] + 13.97}$).}
\label{fig:timewalk}
\end{figure}

To eliminate the effect of time-walk due to leading-edge discrimination, a correction to hit timing was applied using the measured correlation between the signal timing and signal charge, as shown in Fig.~\ref{fig:timewalk}, from where a correction formula was derived.


\section{Results of measurements}
\label{measurements}

\subsection{Time distribution of signals}

\begin{figure}
\centering 
\includegraphics[width=0.8\columnwidth]{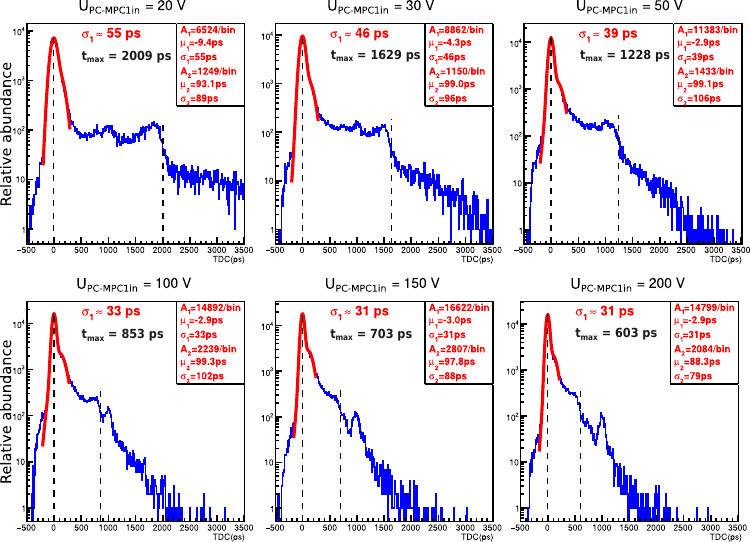}
\caption{The measured timing distribution of pulses in LAPPD \#162 after time walk correction for different potential differences between the photocathode and the microchannel plate 1 ($U_{PC-MCP1in}$). The result of the fit with a sum of two Gaussian functions with parameters $A_i, \mu_i, \sigma_i$ is shown in red, and the fitted parameters are displayed in the box in the top right corner of each of the plots; the dashed lines indicate the peak of the prompt peak and the end of the approximately flat component of the time spectrum ($t_{max}$).
}
\label{fig:timedistr-162}
\end{figure}

\begin{figure}
\centering 
\includegraphics[width=0.8\columnwidth]{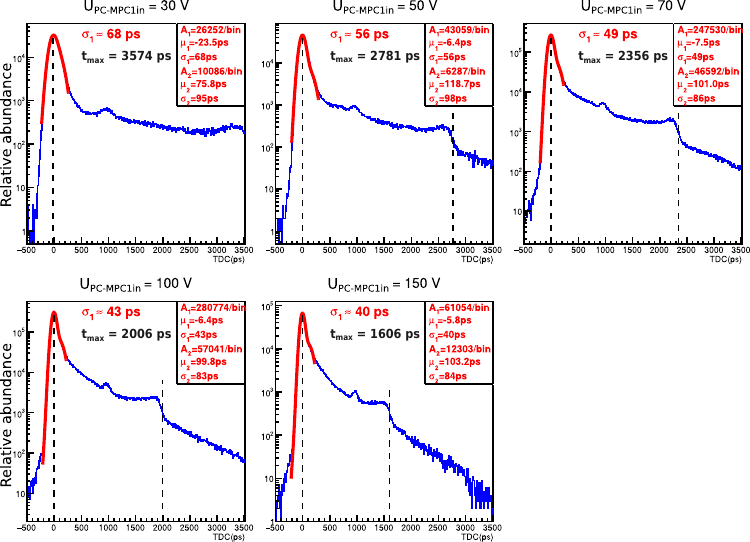}
\caption{
The measured timing distribution of pulses in LAPPD \#109 after time walk correction for different potential differences between the photocathode and the microchannel plate 1 ($U_{PC-MCP1in}$). The result of the fit with a sum of two Gaussian functions with parameters $A_i, \mu_i, \sigma_i$ is shown in red, and the fitted parameters are displayed in the box in the top right corner of each of the plots; the dashed lines indicate the peak of the prompt peak and the end of the approximately flat component of the time spectrum ($t_{max})$.
}
\label{fig:timedistr-109}
\end{figure}

The time spectrum of pulses when corrected for time walk (Figs.~\ref{fig:timedistr-162} and \ref{fig:timedistr-109}) is typical for a MCP-PMT~\cite{korpar-burle-3}. Two distinctive features can be identified: a dominant prompt peak and a uniform distribution of delayed pulses, with a shoulder-like transition region between the two. As will be discussed in Sec.~\ref{sec:model}, the prompt peak comes from photoelectrons that hit the first MCP and start the multiplication, while the uniform distribution of delayed pulses is due to back-scattered photoelectrons that retain most of their kinetic energy. We note here that a full timing distribution can only be measured if the pad side is at least four times the distance between the photocathode and the first MCP, as will become clear in Sec.~\ref{sec:model1}. We also note that a small peak at 1~ns is not affected by varying the potential difference between the photocathode and the microchannel plate 1 ($U_{PC-MCP1}$); it is most likely caused by a reflection\footnote{This explanation is further corroborated by the observation that the width of this spurious peak scales in the same way as the width of the main prompt peak.}. 

In Figs.~\ref{fig:timedistr-162} and \ref{fig:timedistr-109}, we observe that both the width of the dominant peak and the length of the tail decrease with increasing potential difference $U_{PC-MCP1}$, consistent with previous studies of MCP-PMTs~\cite{Lehmann:2024grg,DOLENEC2024169864}. To study this effect systematically, we extract the width of the dominant prompt peak from each of the timing spectra by fitting the initial portion of the distribution with a sum of two Gaussian functions, one representing the prompt peak and the other modeling the shoulder-like transition region, as illustrated in Figs.~\ref{fig:timedistr-162} and \ref{fig:timedistr-109}.

%

%
\begin{figure}
\centering 





\includegraphics[width=0.7\columnwidth]{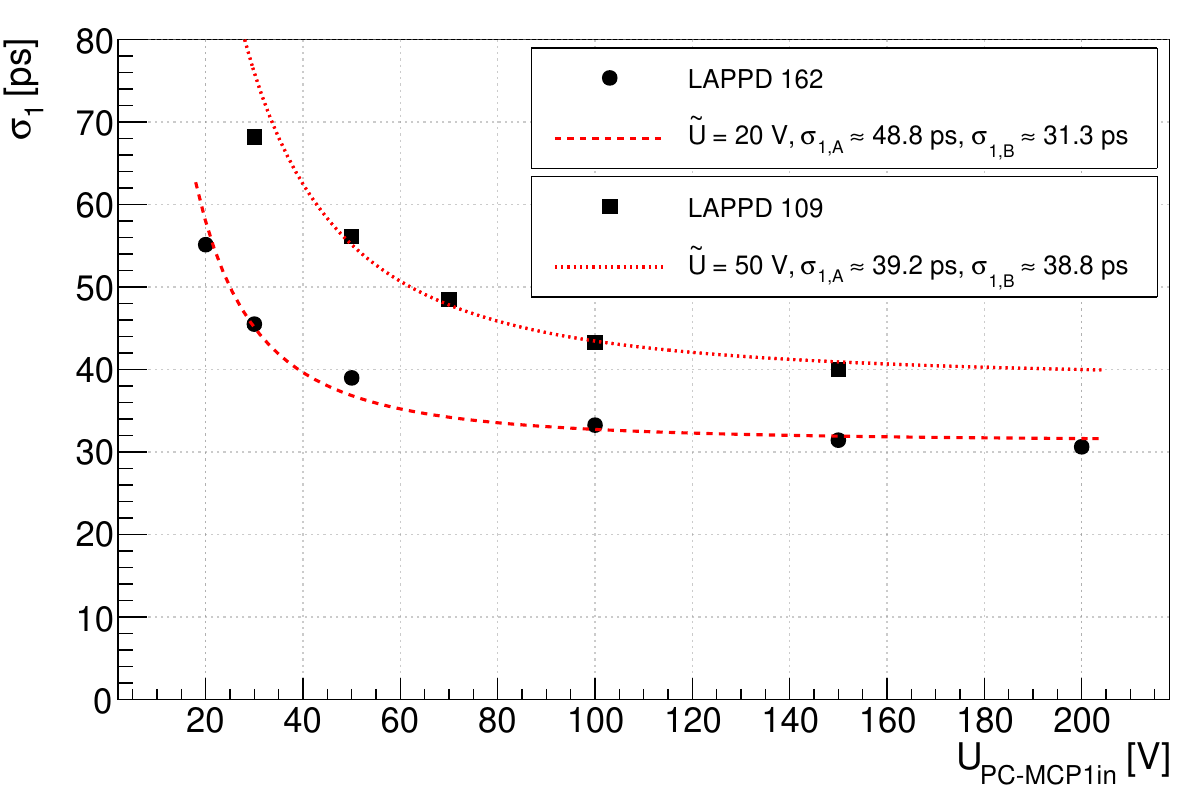}
\caption{Data points: width ($\sigma_1$) of the prompt peak as a function of the potential difference between the photocathode and the microchannel plate 1 ($U_{PC-MCP1in}$) for both samples; statistical errors of the fitted values of $\sigma_1$ are smaller than the marker size. Superimposed: fit to the data points by the model prediction, Eq.~\ref{sigma1-fit}, discussed in Sec.~\ref{sec:model1}. 
}
\label{fig:timing-sigma-u1}
\end{figure}
\begin{figure}
\centering 
\includegraphics[width=0.65\columnwidth]{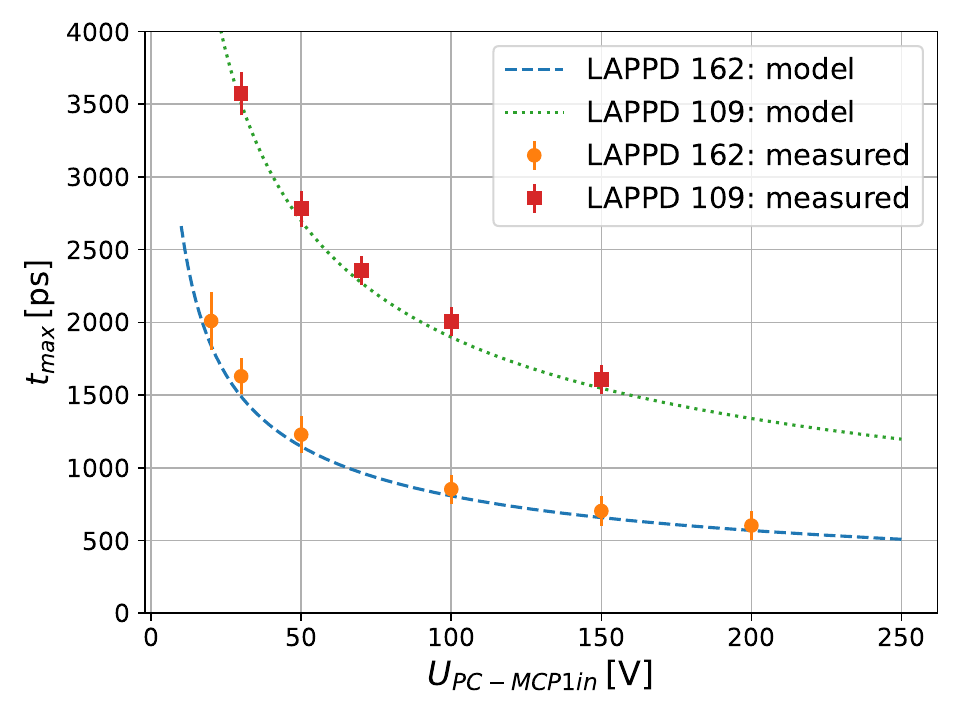}
\caption{The measured width $t_{max}$ of the uniform part of the signal timing distribution, extracted from Figs.~\ref{fig:timedistr-162} and \ref{fig:timedistr-109}, as a function of the potential difference between the photocathode and the microchannel plate 1 ($U_{PC-MCP1in}$); also shown is the model prediction (curves) for  $t_{1,max}$, Eq.~\ref{t1}, as discussed in Sec.~\ref{sec:model1}.
}
\label{fig:timing-uniform-width-u1}
\end{figure}

As shown in Fig.~\ref{fig:timing-sigma-u1}, the time resolution of the prompt peak improves with increasing $U_{PC-MCP1in}$, saturating at $\sigma \approx 30$~ps and $\approx 40$~ps for samples \#162  and  \#109, respectively. Similarly, the width of the uniform distribution is shrinking as a function of $U_{PC-MCP1in}$ as can be seen in Fig.~\ref{fig:timing-uniform-width-u1}; the origin of this dependence will be discussed in Sec.~\ref{sec:model}. 



\subsection{Spatial distribution of induced signal charge}

The capacitive coupling of the anode with the sensing electrode and read-out electronics over a few mm-thick back plate of the LAPPD is bound to influence the signal distribution on the read-out pads due to the spread of induced charge on the coupled electrode. 
\begin{figure}
\includegraphics[width=0.5\columnwidth]{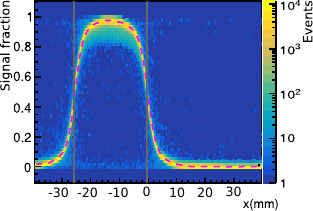}
\includegraphics[width=0.40\columnwidth]{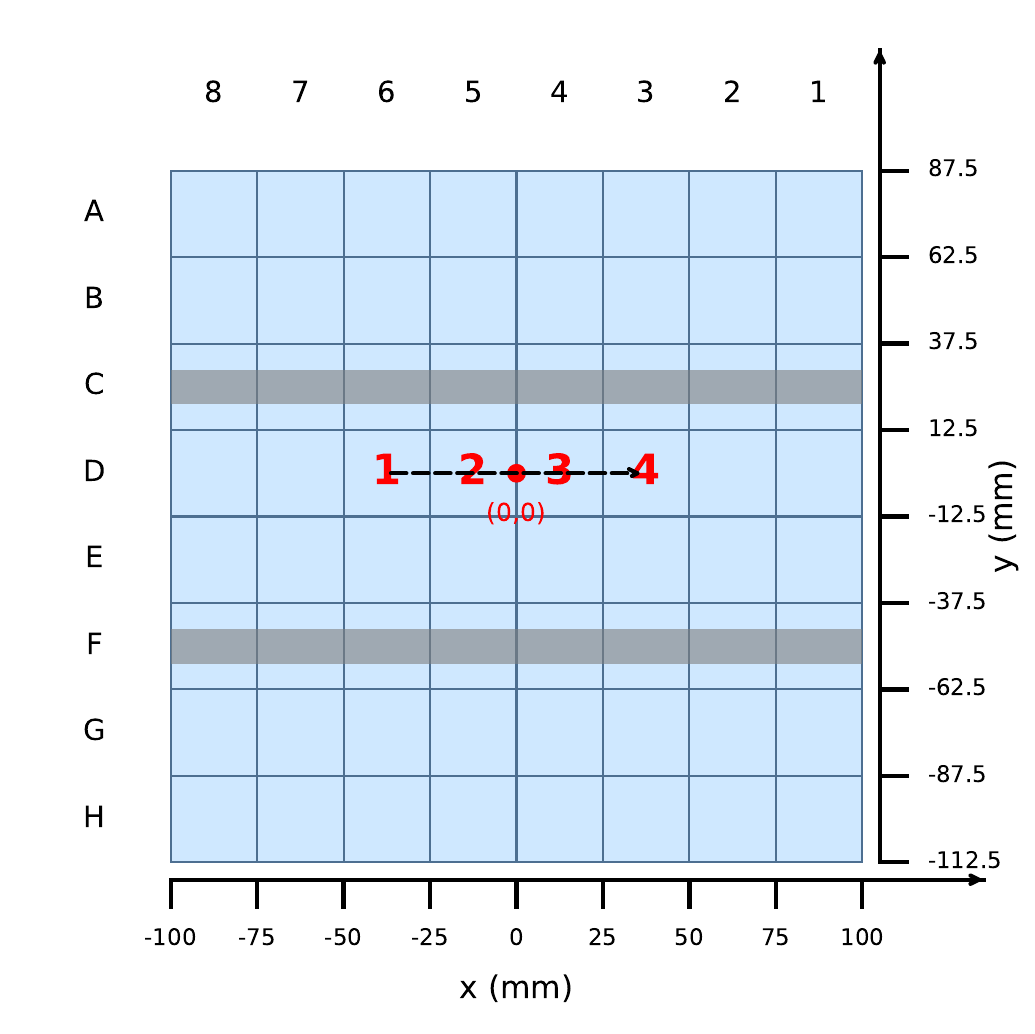}
\caption{Fraction of the signal collected on the reference pad 2 as a function of the position of the laser spot, with the dashed curve showing the average measured values (left). Right: the laser is scanned across pads labeled 1, 2, 3, and 4 in red; the gray lines indicate the positions of spacers.}
\label{fig:signalspread-princ}
\end{figure}

To assess this spread, we have scanned the LAPPD with the laser along the x-direction over pads denoted by 1, 2, 3, and 4 (right pane of Fig.~\ref{fig:signalspread-princ}) while keeping the y-coordinate constant, corresponding to the center of pads. The signal collected on the four pads was measured for different positions of the laser spot. The ratio of the signal on the reference pad (in our case, pad number 2) to the sum of signals on all four pads is shown in the left pane of Fig.~\ref{fig:signalspread-princ} as a function of the laser spot position.


\section{Modeling of the response of the detector}
\label{sec:model}

To understand the response of the LAPPD as presented above, we have modeled the propagation of the photoelectron from the photocathode to the first MCP input plane (MCP1in in Fig~\ref{fig:layout}) as well as the generation of signals by secondary electrons once they exit the multiplication process (MCP2out in Fig~\ref{fig:layout})  and reach the resistive anode. We note that the model is applicable to any MCP-PMT that is thin compared to its lateral dimensions. Such a detector can be approximated by a stack of parallel, infinite layers. Mechanical support structures, such as spacers, are not considered, and the model is of limited value in their immediate vicinity. 

As we demonstrate in the following sections, the propagation model for photoelectrons explains the primary features of the detector's timing response, while the propagation of secondary electrons towards the anode determines the formation of the signal on the sensing electrode. 

\subsection{Modeling of the time response}
\label{sec:model1}

\begin{figure}
\centering 
\includegraphics[width=0.49\columnwidth]{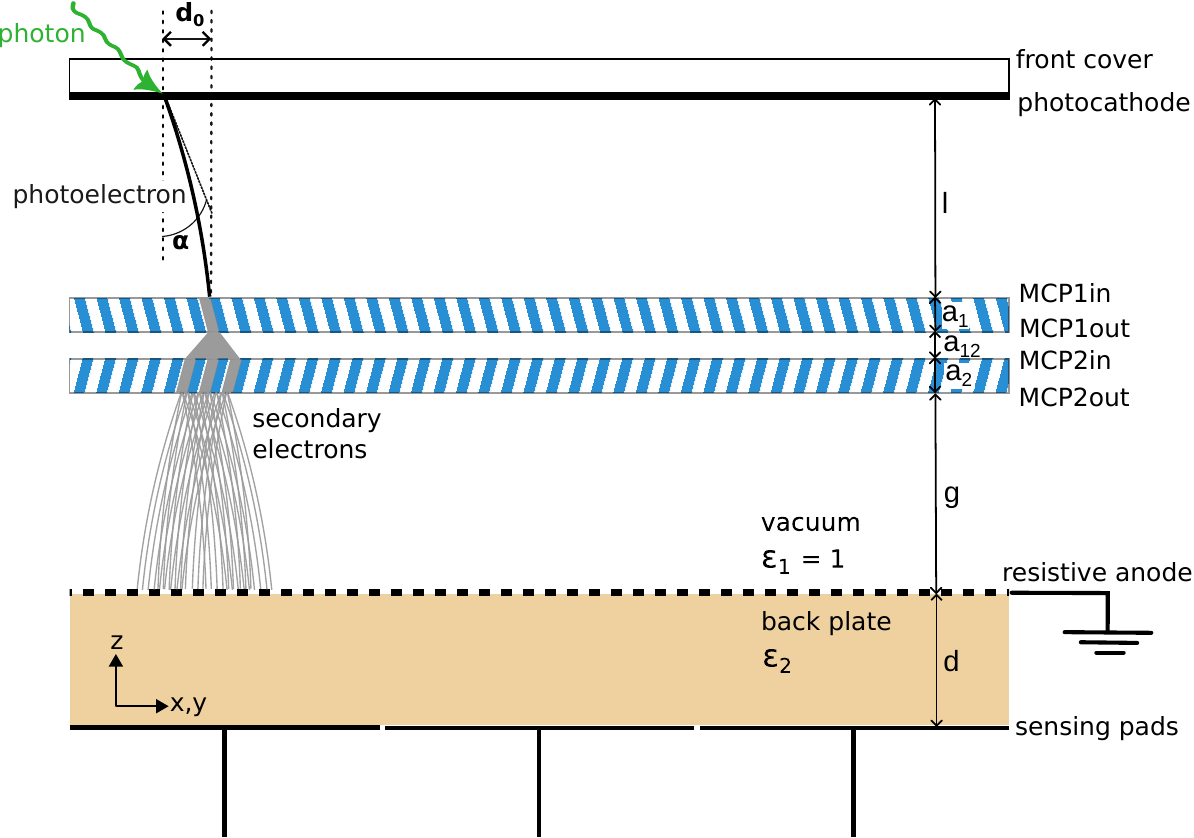}
\includegraphics[width=0.49\columnwidth]{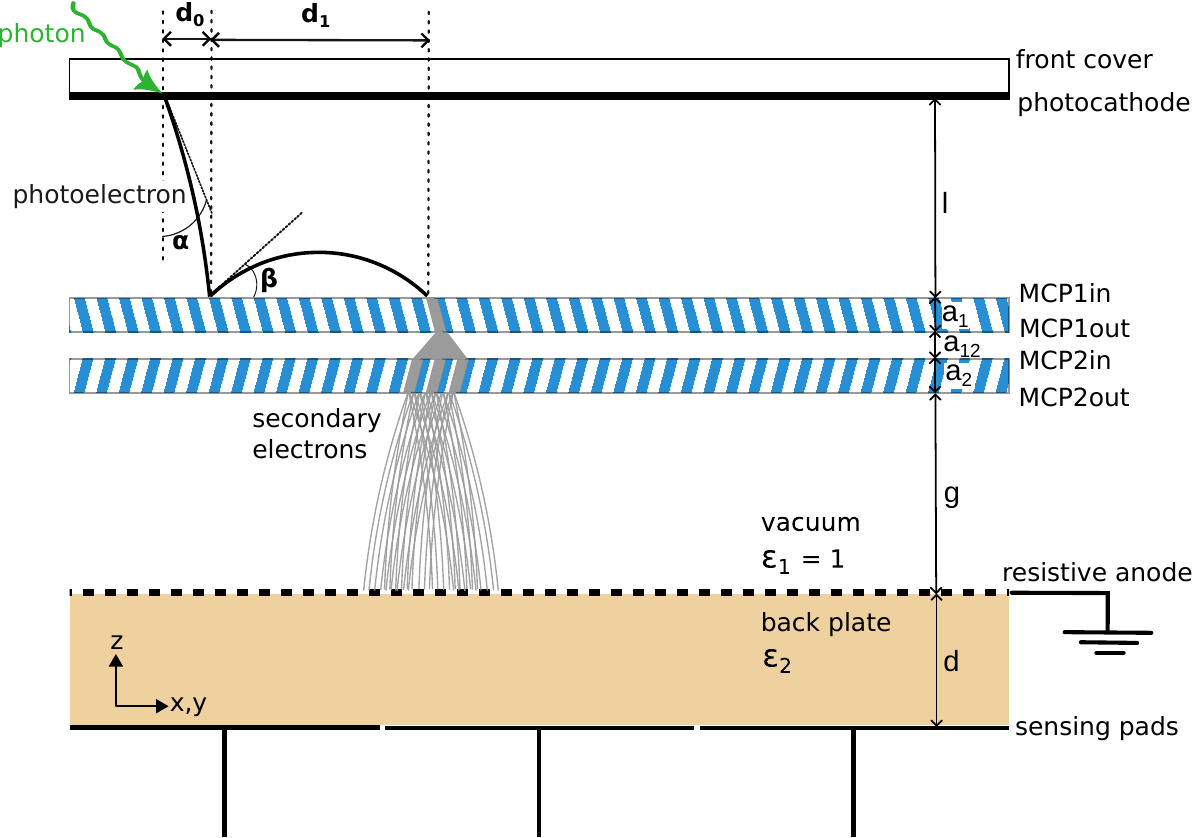}
\caption{Photoelectron trajectories in the MCP-PMT: direct (left) and back-scattered photo-electron (right). 
 }
\label{fig:processes-1}
\end{figure}

When a photon hits the photocathode, it ejects a photoelectron at an angle $\alpha$ relative to the normal as shown in Fig~\ref{fig:processes-1}. 
The trajectory of this electron determines the subsequent development of the signal. Most of the photoelectrons enter a microchannel and undergo multiplication. However, some photoelectrons are reflected from the MCP top surface, either elastically or inelastically; they return to the MCP at a different location, where they trigger multiplication, as shown in Fig.~\ref{fig:processes-1}. 



When analyzing the propagation of a photoelectron from the photocathode to the first MCP (Fig.~\ref{fig:processes-1}, right), we assume a uniform electric field in the gap between these two elements (the width of this gap $l$ is typically a few millimeters). The initial kinetic energy of the photo-electron, $T_0$, which depends on the photon energy and the work function of the photocathode material, is of the order of 1~eV,
and is thus small compared to the energy gained from the electric field, $T=e_0U_{PC-MCP1in}$; in the following derivations, we will, therefore, only retain the leading term in the $T_0/T$ expansion.  

Using straightforward kinematic relations, analogous to the kinematics
of a projectile fired at an oblique angle, we arrive at the following relations  for the maximal lateral deflection $d_0$ of the photoelectron when hitting the first MCP and the maximal time of flight $t_0$, 
\begin{equation}
   d_0 = 2 l \sqrt{\frac{T_0}{T}}\>, 
\end{equation}
\begin{equation}
   t_0 = l \sqrt{\frac{2m_e}{T}}  \>.
    \label{t0}
\end{equation}
The results show that both quantities depend on the applied voltage. Higher voltage and smaller gap reduce both the travel time and the lateral spread, which is desirable for precise timing and spatial resolution. The difference $\Delta t_0$ of the largest and smallest $t_0$, i.e., for photoelectrons that are emitted from the photocathode at the angles of $\alpha=90^{\circ}$ and $\alpha=0^{\circ}$, respectively,
\begin{equation}
\label{eq:delta-t0}
\Delta t_0 = 
\frac{l\sqrt{2m_e T_0}}{T}=\frac{l\sqrt{2m_e T_0}}{e_0U_{PC-MCP1in}} ,
\end{equation}
is a measure of the time resolution of the device.
For one of the LAPPD samples studied (\#162), the numerical values are as follows. At $U_{PC-MCP1in}=30$~V, the maximal flight time to the first MCP is 400~ps, the time spread is $\Delta t_0 = 200$~ps, and the maximum lateral deflection $d_0$ amounts to 0.24~mm; for $U_{PC-MCP1in}=100$~V, the correspoding values are 220~ps, 60~ps, and 0.13~mm. Here we assume a value of $T_0=1$~eV for the initial kinetic energy of the photoelectron, taking into account the work function of a multi-alkali photocathode and the photon energy (3.06~eV at the wavelength of 405~nm). 

The time spread $\Delta t_0$ depends on the potential difference between the photocathode and the first MCP, $U_{PC-MCP1in}$, and varies as $1/U_{PC-MCP1in}$. This spread 
is related to the width of the main component of the time distribution of single photon signals, the dominant Gaussian-like contribution shown in Figs.~\ref{fig:timedistr-162} and \ref{fig:timedistr-109}. As shown in Fig.~\ref{fig:timing-sigma-u1}, the measured width of the prompt peak ($\sigma$ of the gaussian-like contribution, denoted as $\sigma_1$), is indeed reduced by increasing $U_{PC-MCP1in}$; it eventually saturates because of other sources of uncertainty (laser, electronics, time evolution in the multiplication process). A fit of a quadratic sum of the term that varies as $1/U_{PC-MCP1in}$ and a constant contribution, 
\begin{equation}
    \sigma_1=\sqrt{\sigma_{1,A}^2 ( \frac{\tilde{U}}{U_{PC-MCP1in}} )^2 +\sigma_{1,B}^2 }  \label{sigma1-fit}
\end{equation}
is displayed as a dashed curve in Fig.~\ref{fig:timing-sigma-u1}; it shows a good agreement with the data points. Here $\sigma_{1,A}$ and $\sigma_{1,B}$ are free parameters, and $\sigma_{1,A}$ is the value of the first term at $U_{PC-MCP1in}=\tilde{U}$. 

The term  $\sigma_{1,A} ( \frac{\tilde{U}}{U_{PC-MCP1in}} )$ which we will denote as $\sigma_A$, is the component of the resolution due to photoelectron propagation. Its value at a given value of $U_{PC-MCP1in}$ is proportional to $\Delta t_0$ (Eq.~\ref{eq:delta-t0}), $\sigma_A = k \Delta t_0$. For sample \#162, the value of this coefficient is $k=0.24$, while for  \#109, the corresponding value is $k=0.21$. It is interesting to observe that these values are in good agreement with the prediction of a model discussed in \cite{Korpar:2008ria}.

The difference in the saturated value of $\sigma_1$ for the two samples is probably due to the difference in time evolution in the multiplication process in the micro-channels with different diameters (20~$\mu$m for \#109, 10~$\mu$m for \#162). Since both samples have the same length-to-diameter ratio, the electric field in the 20~$\mu$m micro-channel is lower than in the 10~$\mu$m case, so that the multiplication proceeds at a lower rate, which would lead to a different contribution to the overall time resolution.

The photoelectrons striking the MCP do not always enter a micro-channel but may instead scatter from its surface, as has already been discussed in previous studies of MCP-PMTs~\cite{Korpar:2008ria,korpar-burle-3}. If the scattering is elastic, the electron maintains its kinetic energy but is deflected by an angle $\beta$ relative to the surface. Depending on the scattering angle, the electron can re-enter the MCP at a different position with a separation of
\begin{equation}
    d_1=2l\sin(2\beta) \>, \label{d1}
\end{equation}
from the original impact point on the MCP. The time of arrival of the scattered photoelectron back to the MCP is 
\begin{equation}
    t_1=2l\sin\beta \sqrt{\frac{2m_e}{T}} = 2l\sin\beta \sqrt{\frac{2m_e}{e_0U_{PC-MCP1in}}} 
    \label{t1}
\end{equation}
with a maximal value of $t_{1,{\rm{max}}} = 2 t_0$.
The maximum range is equal to two times the photocathode-to-MCP1 distance $d_{1,{\rm{max}}} = 2 \ l$.  

These results indicate that the scattering distance exceeds several millimeters (for sample \#162, the maximum range is 2.38~mm), significantly larger than the direct deflection $d_0$ calculated above, effectively displacing the impact point of the photoelectrons. 

Here we introduce a probabilistic model, assuming an angular distribution uniform over the solid angle of the elastically scattered photo-electrons.
This enables the calculation of the spatial distribution for the landing positions of elastically backscattered photoelectrons, $d^2P/dxdy$. 
%
\begin{figure}
\centering 
\includegraphics[width=0.6\columnwidth]{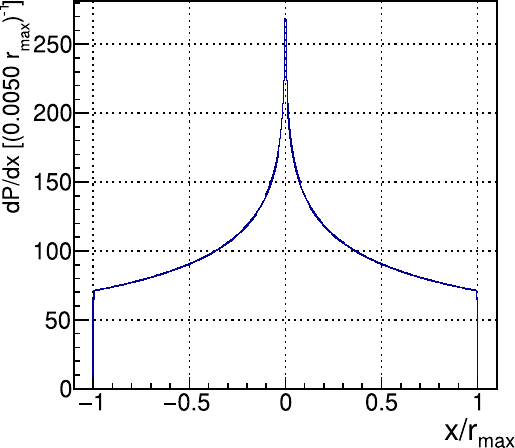}
\caption{Elastically back-scattered photoelectrons: distribution of impact points in one projection, with $r_{\rm{max}} = d_{1,{\rm{max}}} = 2 \ l$.
}
\label{fig:backscattered-photoelectrons}
\end{figure}
The resulting projection of the spatial distribution is displayed in Fig.~\ref{fig:backscattered-photoelectrons}. 

The time-of-flight of the elastically-scattered photo-electrons is uniformly distributed between $t_1=0$ and $t_1=2 t_0$ as can be easily seen following a straightforward analytical derivation\footnote{The distribution over $t_1$, $dN/dt_1$, can be expressed as $dN/dt_1 = dN/d(\sin{\beta}) \cdot d(\sin{\beta})/dt_1$; this is constant since $d(\sin{\beta})/dt_1$ is a constant according to Eq.~\ref{t1}, and $dN/d(\sin{\beta})$ is a constant due to our assumption that the angular distribution of scattered photoelectrons is uniform over the solid angle.}. Such a uniform distribution in the tail of the timing response has already been reported in our earlier studies of MCP-PMTs~\cite{Korpar:2008ria,korpar-burle-3}, and is confirmed in the present study (Figs.~\ref{fig:timedistr-162} and \ref{fig:timedistr-109}).

Further support for the model is provided by Fig.~\ref{fig:timing-uniform-width-u1}, which shows reasonably good agreement between the measured width of the uniform component of the signal-timing distribution and the model prediction as a function of the potential difference between the photocathode and microchannel plate 1 ($U_{PC\text{–}MCP1in}$). The prediction is based on the expression for $t_{1,\rm{max}}$, the maximum time required for elastically scattered photoelectrons to reach MCP1, as given in Eq.~\ref{t1}. The small systematic discrepancy between the measurements and the model warrants further investigation; potential contributing factors include secondary scattering of elastically scattered photoelectrons.

These results establish that back-scattering of photoelectrons is a significant contributor to the spread of the induced signal, both in time and in space. It also shows that the flat part of the distribution in time, as shown in Figs.~\ref{fig:timedistr-162} and \ref{fig:timedistr-109}, can be attributed to elastic back-scattering of photoelectrons.

\subsection{Modeling of the spatial distribution of induced signal charge}
\label{sec:model2}

In modeling the signal generated by secondary electrons, we introduce a simplifying assumption: the distribution of secondary electrons during their transit from the exit of the second MCP to the resistive anode (Fig.~\ref{fig:processes-1}) is treated as a point-like charge. Their initial kinetic energy is assumed to be negligible, such that the dominant contribution arises from the induced charge on the segmented sensing-electrode structure.
Later in this section, we will present a special measurement that verifies this assumption. Further support for this assumption comes from the literature~\cite{zero-T0-1,zero-T0-2,zero-T0-3} and our previous studies of MCP-PMTs with internal segmented anodes~\cite{korpar-burle-3}, where the transition in measured signal charge on the reference pad is very sharp when scanning with a laser across the boundary between the pads. We also assume that the charge arriving at the resistive anode does not significantly drift during the period of signal development.

The Shockley–Ramo theorem~\cite{ramo,Shockley} provides the foundation for calculating the induced charge on a given pad of the sensing electrode due to charges at positions $\vec{r_i}(t)$. The induced charge at the electrode $j$ is at time $t$ 
\begin{equation}
\label{SR}
 Q_j(t) = - \sum_{i=1}^N q_i u_j(\vec{r_i}(t))
\end{equation}
where the weighting field $u_j$ satisfies the equation
\begin{equation} 
\nabla \left[ \epsilon(\vec{r}) \nabla u_j(\vec{r})\right] = 0 
\end{equation}
with appropriate boundary conditions at the electrodes. Denoting by $Q_{\rm anode}$ the charge carried by secondary electrons that landed at the resistive anode, the charge induced at the sensing electrode is given by the following expression
\begin{equation}
\label{eq:qanode-qsignal}
Q_{\rm sens.el.} = - Q_{\rm anode} \ \left( 1+\frac{\epsilon_1}{\epsilon_2}\frac{d}{g} \right)^{-1}.
\end{equation}
The charge collected at a single pad of the sensing electrode at $z=0$ amounts to
\begin{equation}
Q_{\rm pad} (x,y,z=d) = - Q_{\rm anode} \ u(x,y,z=d),
\end{equation}
where $u(x,y,z=d)$ refers to the specific pad, and is evaluated numerically.
Here, as in Eq.~\ref{SR}, the coordinates $x, y, z=d$ refer to the coordinates of the secondary electrons that have reached the resistive anode (at $z=d$).
The charge induced on a single pad is then normalized against the total charge induced on the sensing electrode, giving a fractional signal that can be compared to measurements.

\begin{figure}
\includegraphics[width=0.95\columnwidth]{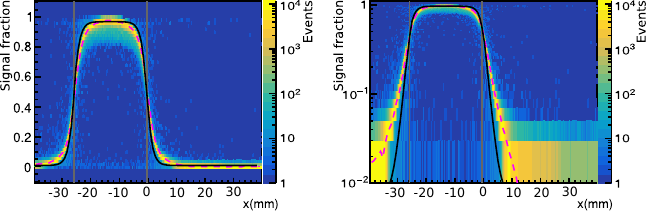}
\caption{Fraction of the signal charge collected on the reference pad 2, $Q_2/(Q_1+Q_2+Q_3+Q_4)$, versus the position of the laser spot relative to the center of the reference pad for sample \#162.  The full black curve represents the model's prediction without back-scattering of secondary electrons from the resistive anode, and the dashed curve shows the average measured values. 
}
\label{fig:signalspread-162-0}
\end{figure}

To compare our model predictions for charge sharing with the measured data, we restrict the discussion to the main component — that is, events in which the photoelectron promptly initiates multiplication in microchannel plate 1. In the experimental data, this selection is implemented by choosing events within the main prompt peaks of Figs.~\ref{fig:timedistr-162} and \ref{fig:timedistr-109}.
The signal spread observed in Fig.~\ref{fig:signalspread-162-0} agrees reasonably well with the prediction of the model. However, the observed tails are substantially more pronounced than the model anticipates, and the corresponding transitions appear noticeably less sharp.

\begin{figure}
\centering 
\includegraphics[width=0.85\columnwidth]{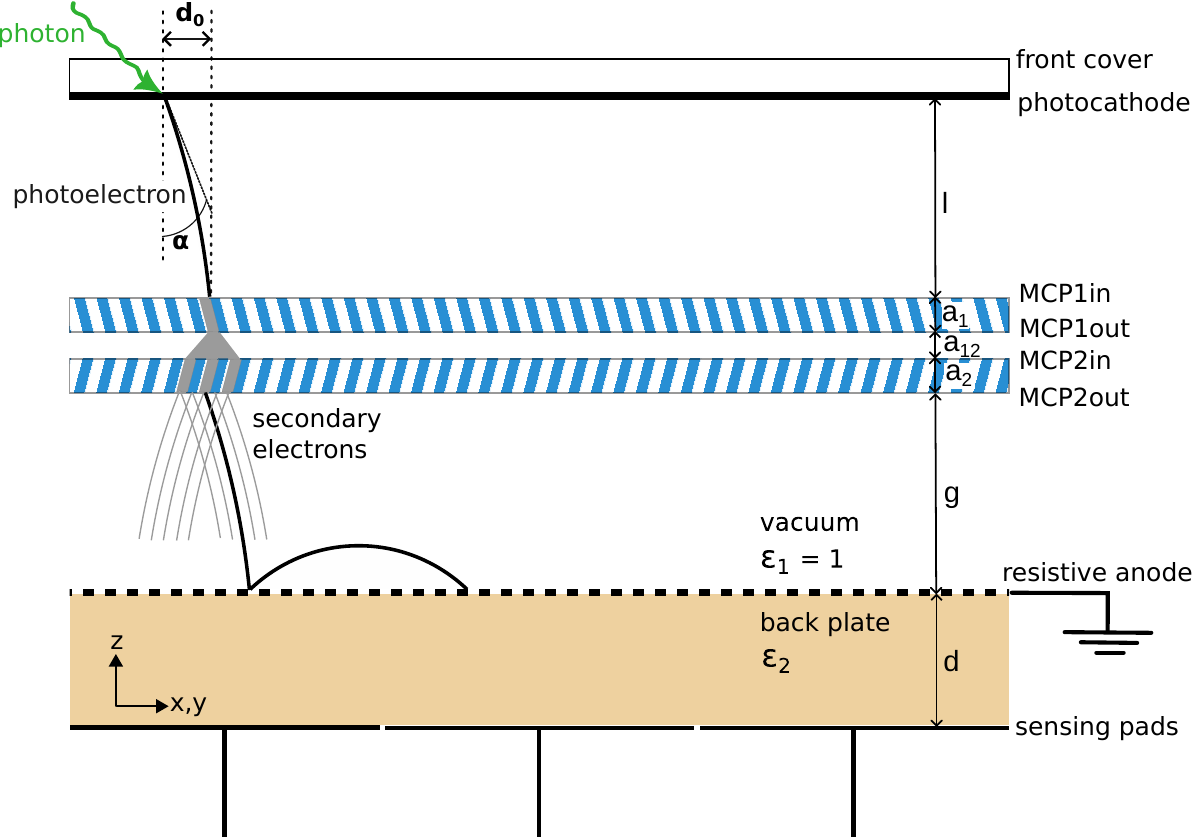}
\caption{Schematic depiction of the propagation of the secondary electrons during signal development.}
\label{fig:processes-2}
\end{figure}

To account for this discrepancy, we note that low-energy electrons of kinetic energy of a few hundred eV that impinge on a solid surface of the resistive anode can get backscattered with a sizable probability, between 10\% and 30\%~\cite{backscattering-anode-literature}, or emit secondary electrons~\cite{LAPINGTON1997336}. Such back-scattering could lead to the process shown in Fig.~\ref{fig:processes-2}, i.e., a fraction of secondary electrons that have reached the anode would scatter and reach the anode again, but at a displaced position.
We therefore refine our model and assume that a fraction of the secondary electrons, denoted by $\eta$, stops immediately at the resistive anode, while the remaining portion $(1 - \eta)$  elastically scatters before being absorbed on the anode. The scattering behavior is modeled using the distribution derived above (Sec.~\ref{sec:model1}), with the maximum scattering range (Eq.~\ref{d1}) reaching twice the size of the gap between the second micro-channel plate and the resistive anode (6~mm for sample \#162 and 13.4~mm for \#109). 

The resulting charge collected at a pad of the sensing (read-out) electrode is now expressed as a weighted sum of two terms: one corresponding to the direct, non-scattered electrons, and the other representing the convolution 
\begin{equation}
\label{eq-convol}
\left( \frac{d^2P}{dx dy}*u(z=d)\right) (x,y) 
\end{equation}
of the weighting potential of the pad with the spatial distribution of scattered electrons,
\begin{equation}
\label{eq-pixel-q}
Q_{\rm pad} (x,y,d) = - Q_{\rm anode} \ \left[ \eta \ u(x,y,d) + (1-\eta)  \left( \frac{d^2P}{dx dy}*u(z=d)\right) (x,y) \right]   
\end{equation}

\begin{figure}
\centering 
\includegraphics[width=0.7\columnwidth]{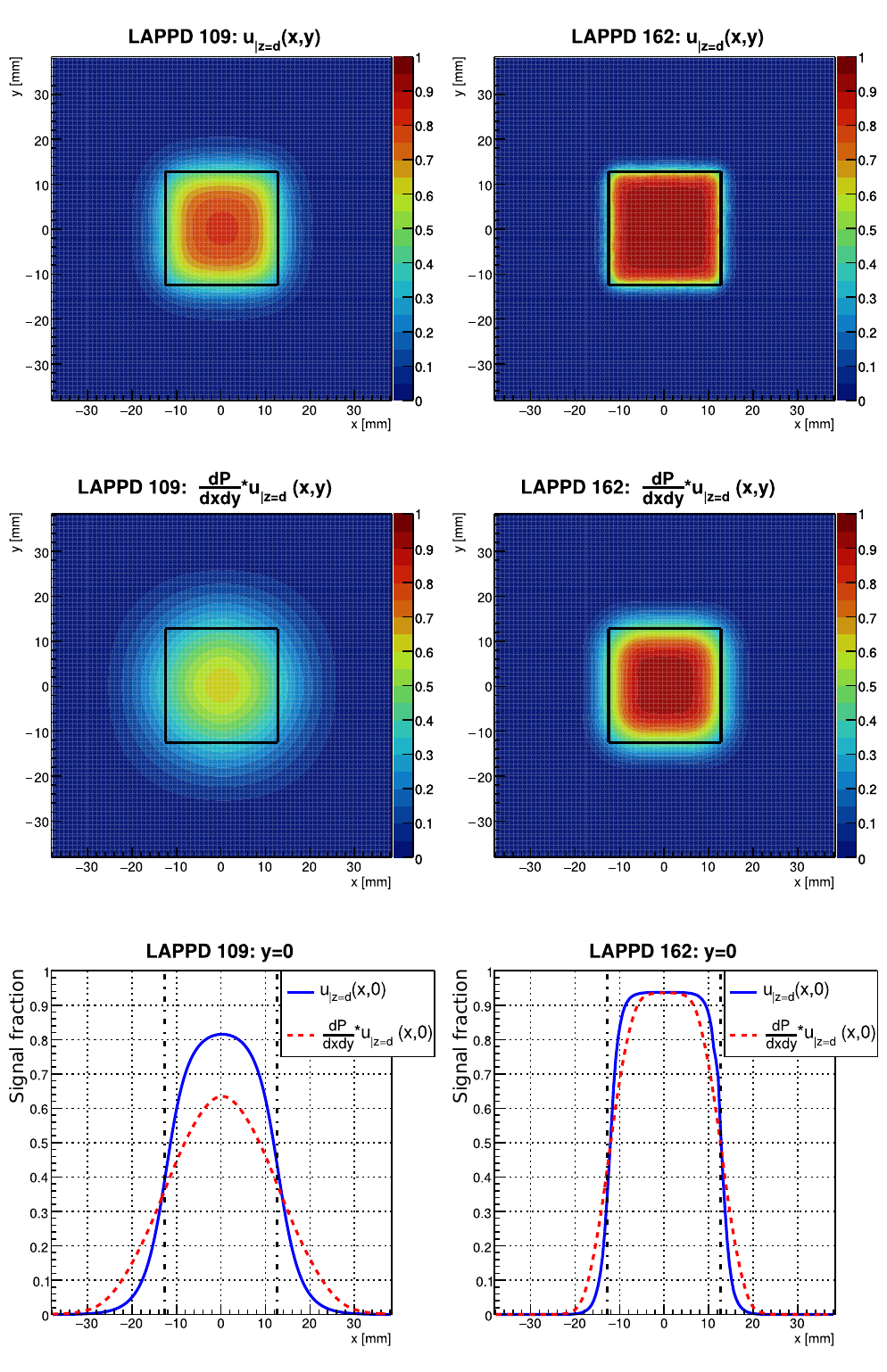}
\caption{Two components of the induced charge distribution for LAPPD geometries of samples \#109 (left) and \#162 (right) with 25.4~mm pads: without (top) and with secondary electron back-scattering (middle); bottom: charge sharing profile as a function of x for $y=0$. 
}
\label{fig:charge-sharing-25.4mm}
\end{figure}

\begin{figure}
\includegraphics[width=0.95\columnwidth]{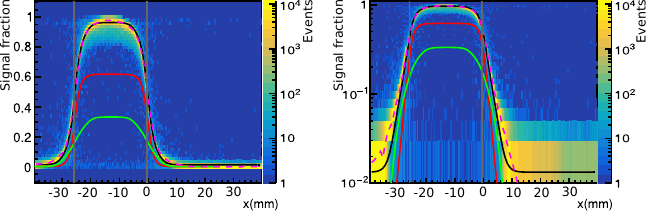}
\caption{Fraction of the signal charge collected on the reference pad 2, $Q_2/(Q_1+Q_2+Q_3+Q_4)$, versus the position of the laser spot relative to the center of the reference pad for sample \#162.  The dashed curve represents the measured values, while the black curve represents the model fitted to the data, a sum of unscattered (red) and scattered (green) contributions; the gray lines indicate the pad boundaries. 
}
\label{fig:signalspread-162}
\end{figure}

\begin{figure}
\includegraphics[width=0.95\columnwidth]{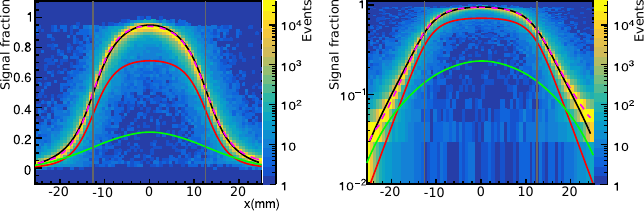} 
\caption{Same as Fig.~\ref{fig:signalspread-162} for sample \#109.  
}
\label{fig:signalspread-109}
\end{figure}

The predictions of the two components of the revised model are shown in Fig.~\ref{fig:charge-sharing-25.4mm} for both LAPPD samples. 
This division into directly absorbed and scattered electrons allows for a more realistic description of how induced charge is spread across the sensing electrode, as can be seen from Figs.~\ref{fig:signalspread-162} and \ref{fig:signalspread-109}. The sum of the two components (Eq.~\ref{eq-pixel-q}) is fit to the data by varying the parameter $\eta$.
It is interesting to note that the value of $(1-\eta)$, i.e., the probability of scattering as returned by the fit, amounts to 30\% and 35\% for the two samples, in good agreement with the value expected from the literature~\cite{backscattering-anode-literature}. 


Finally, we performed a dedicated measurement to verify our assumption that the swarm of secondary electrons at the exit of the second MCP can be considered as a point-like charge. This check was carried out by comparing the charge sharing at two different potential differences between microchannel plate 2 and resistive anode ($U_{MCP2out-A}$), 500~V and 100~V.  
\begin{figure}
\centering 
\includegraphics[width=0.95\columnwidth]{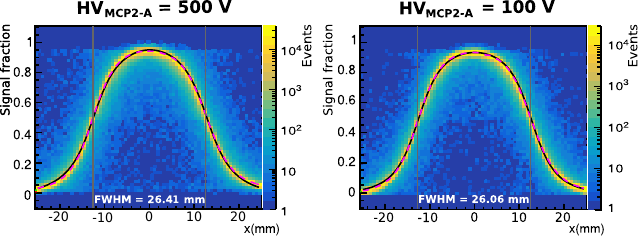}
\caption{Fraction of the signal charge collected on the reference pad 2, $Q_2/(Q_1+Q_2+Q_3+Q_4)$, versus the position of the laser spot relative to the center of the reference pad for $U_{MCP2out-A} = 500$~V (left) and $U_{MCP2out-A} = 100$~V (right).  The dashed curve represents the measured values, while the black curve represents the model fitted to the data; the gray lines indicate the boundaries of the pad. Sample \#109 was used in this measurement. 
}
\label{fig:signalspread-u5}
\end{figure}
We again perform scans with the laser spot moving from the centre of the left neighboring pad to the centre of the right neighboring pad for the two $U_{MCP2out-A}$ settings. The comparison between the results, shown in Fig.~\ref{fig:signalspread-u5}, shows an almost identical charge sharing, pointing to a negligible influence of the value of $U_{MCP2out-A}$. From this observation, we conclude that the by far dominant contribution of charge sharing is the induced charge spread on the sensing electrode, whereas the contribution of initial secondary electron spread can be neglected, thus confirming the justification for the model assumption. This result is also in good agreement with our previous studies of MCP-PMTs with internal segmented anodes~\cite{korpar-burle-3}, mentioned above.

\subsubsection{Signal development for different detector parameters}
\label{sec:model3}

Having observed the success of our model in reproducing features observed in the bench tests of two LAPPD samples with very different geometric and material properties, we are now able to predict the performance of MCP-PMTs with capacitively coupled sensing electrodes in other geometries and with other material properties. 
To investigate the dependence, we have carried out modeling of signal development for different parameters of the MCP-PMT, the distance from MCP2 to the resistive anode $g$, the thickness $d$, and the dielectric constant $\epsilon_2$ of the back plate. 

\begin{figure}
\centering 
\includegraphics[width=0.8\columnwidth]{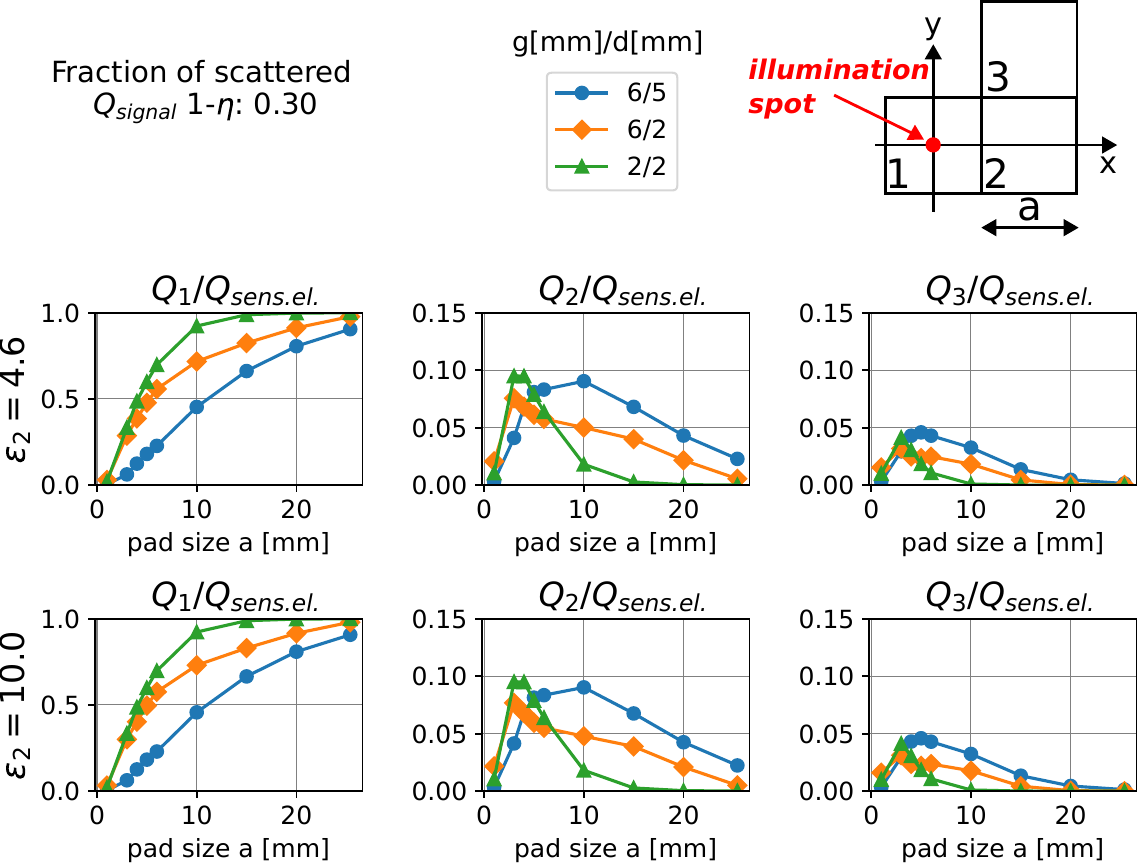}
\caption{Induced charge distribution on a single pad, normalized to the charge induced at the sensing electrode as a function of pad size for different dielectric constants $\epsilon_2$ of the back plate material and for different gap-to-resistive anode thickness ($g/d$) ratios: for the pad bellow the photon impact point (top), the next-neighbor pad (middle), and the diagonal-neighbor pad (bottom). The calculation assumes a 30\% probability for back-scattering of secondary electrons off the resistive anode.}
\label{fig:charge-sharing-scatt}
\end{figure}

The results are presented in Fig.~\ref{fig:charge-sharing-scatt} and Table~\ref{tab:qanode-qsignal}. As expected, the degree of charge sharing between neighboring pads exhibits a strong dependence on the detector geometry.
%
%
\begin{figure}
\centering 
\includegraphics[width=0.8\columnwidth]{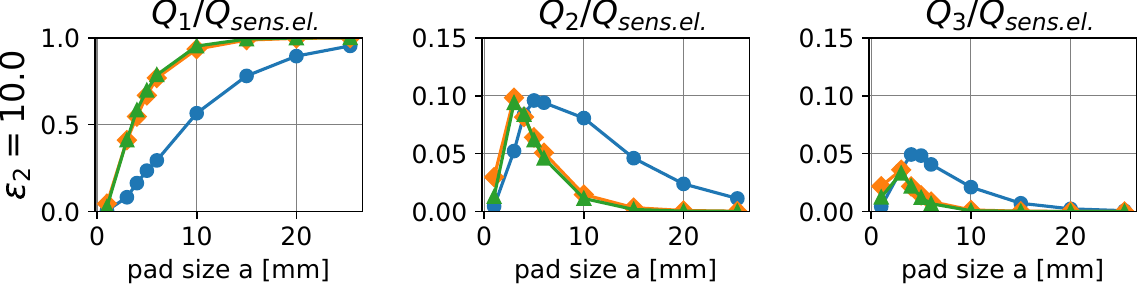}
\caption{Same as Fig.~\ref{fig:charge-sharing-scatt}, but assuming no back-scattering of secondary electrons from the resistive anode.}
\label{fig:charge-sharing-no-scatt}
\end{figure}
For relatively small pads (smaller than 5~mm), more than 50\% of the charge is induced on neighboring pads (Fig.~\ref{fig:charge-sharing-scatt}). Such effects become negligible for large pads. This outcome demonstrates that the segmentation of the sensing electrode must be carefully matched with the intended application of the detector: a fine segmentation provides high spatial resolution, but at the cost of increased charge sharing. However, the value of the dielectric constant of the back plate material has only a minor effect on the charge sharing.

Figure~\ref{fig:charge-sharing-no-scatt} shows the charge-sharing behavior in the absence of backscattering from the resistive anode. We note that this scenario also effectively describes the MCP-PMT operation in a magnetic field of around 1~T, oriented perpendicular to the entrance window — a configuration encountered in some particle physics experiments. In this case, the backscattered electrons spiral along the magnetic field lines and end up at the anode very close to where they originated instead of flying sideways as in Fig.~\ref{fig:processes-2}. Such a magnetic field also has a similar effect on backscattered photoelectrons, as observed in previous studies~\cite{KORPAR2011162,Rieke:2016hqe}.  

Another important factor is the completeness of signal capture, as given by Eq.~\ref{eq:qanode-qsignal}, and shown in Table \ref{tab:qanode-qsignal} for a few examples.
\begin{table}
\caption{Fraction of the signal induced on the sensing electrode for various values of the MCP2-to-resistive-anode distance $g$, the thickness $b$ and the dielectric constant $\epsilon$ of the back plate of the MCP-PMT).}
\label{tab:qanode-qsignal}
\begin{center}
\begin{tabular}{|l|c|c|c|}
\hline
 $g$ distance & $d$ thickness & dielectric constant $\epsilon_2$ &  $-
 Q_{\rm sens.el.}/Q_{\rm anode}$ \\
MCP2-anode & back plate &  of the back plate & \\
\hline
2~mm & 2~mm & 10  & 0.9 \\
\hline
6~mm & 2~mm & 10 & 0.97 \\
\hline
6~mm & 2~mm & 4.6 & 0.93 \\
\hline
6~mm & 5~mm & 4.6 & 0.85 \\
\hline
\end{tabular}
\end{center}
\end{table}
Related to this, it is worth noting that a finite pad array is often used in experiments, so that not all scattered electrons are necessarily collected. In case the total charge information is required, a correction for incomplete charge detection must be applied using Eq.~\ref{eq-pixel-q} and scaling the measured charge fractions accordingly.

\section{Discussion}

The above investigation into the influence of structural parameters, such as the distance between the microchannel plates and the anode, as well as the dielectric properties of the medium between the resistive anode and the segmented sensing electrode (Fig.~\ref{fig:charge-sharing-scatt}), can be summarized as follows. Larger distances or higher dielectric constants both affect the weighting potential, leading to different spreading patterns of the induced charge. These dependencies emphasize that not only pad geometry but also overall detector design plays a crucial role in determining signal characteristics. 

There are several consequences of the results of this study for applications of LAPPDs (and MCP-PMTs in general) with capacitively coupled sensing electrodes in RICH detectors and in PET and TOF-PET devices.   

For a RICH detector, the relevant figure of merit is the resolution in the photon impact point position. A capacitively coupled sensing electrode with center-of-gravity estimation of the impact point position can improve the impact point precision as compared to the MCP-PMTs with internally segmented anodes. In this case, during reconstruction, care must be taken to properly combine hits into clusters, applying methods that have already been developed, for example, for the RICH detector of the ALICE experiment~\cite{alice-rich-reco}. This strategy is considered one of the options for upgrading ARICH, the Cherenkov aerogel counter of the Belle II experiment~\cite{belle2-upgrade}. 
If, however, the track density in the RICH detector reaches the level anticipated for the LHCb RICH, the resulting high occupancy requires pad sizes of roughly 1~mm, together with a timing resolution better than 100~ps for single photons~\cite{LHCb:2021glh}. While the excellent timing performance of the LAPPD makes it a highly promising option, in a capacitively coupled MCP-PMT with such small pads, a single photon hit would distribute charge over a large number of pads. Additional studies are therefore needed, incorporating the results of the present work into simulations of the RICH detector response and into the event-reconstruction algorithms.

For PET or TOF-PET applications, where scintillating crystals are attached to the front window of the LAPPD, the impact points of scintillating photons are spread over several pads. Charge sharing could indeed be beneficial in further improving the resolution of the annihilation gamma conversion point position. The excellent timing properties and large size make LAPPD a very promising candidate for flexible panel-based TOF-PET systems~\cite{razdevsek-panel-tofpet}.  

The present study also reinforces our earlier conclusion regarding the critical role of a thin gap between the photocathode and the first MCP. Reducing this gap improves the time resolution by narrowing the main Gaussian peak and shortening the photoelectron back-scattering tail in the time spectrum. The back-scattering range, approximately twice the gap size, is likewise reduced. Increasing the voltage between the photocathode and the first MCP yields a similar improvement in timing performance.

Finally, it is worth noting that previous studies have reported a reduction in back-scattering effects — whether from primary photoelectrons or from secondary electrons — on the spatial distribution of charge at the sensing electrodes when a magnetic field is applied perpendicular to the MCP-PMT entrance window~\cite{KORPAR2011162,Rieke:2016hqe}. This behavior can be also understood within the framework of the model presented in this paper as discussed in Sec.~\ref{sec:model3}.

\section{Summary and outlook}

 We studied a large-area MCP-PMT, the LAPPD Gen-II version with a resistive anode, featuring borosilicate glass or ceramic back plates, and a capacitively coupled sensing (readout) electrode. We measured the timing response and the induced charge distribution on the read-out electrodes. 
 
 We developed models for photoelectron propagation and signal evolution, and compared them with the measurements. We found a very good agreement between the two and were able to make predictions for performance for different geometries and materials of the back plate of the MCP-PMT.

In future work, we plan to study the performance of LAPPD Gen-II sensors equipped with custom-designed sensing electrodes featuring various segmentation patterns. We also aim to investigate the intermediate region of the timing spectrum, which is currently attributed to inelastic scattering and secondary emission. Additionally, we will investigate how to incorporate the results of this study into simulations of detector response for RICH and TOF-PET applications, as well as how to integrate them into event reconstruction procedures.


\section{Acknowledgements}

We acknowledge very useful discussions with  Incom Inc. staff, Mark A.~Popecki, Alexey Lyashenko, and Cole Hamel. 
This work was supported by the following funding sources: European Research Council, Horizon 2020
ERC Advanced Grant No. 884719, ERC Proof-of-Concept Grant No. 101113474, and Slovenian Research and Innovation Agency (ARIS) research grants No. J1-9124, J1-4358, N1-0422 and P1-0135.

\bibliography{biblio}

\end{document}